\documentstyle[aps,multicol,epsf]{revtex}

\def\beginwide{
        \end{multicols} \vspace*{-0.5cm} \noindent
        \rule{3.5in}{.1mm}\rule{.1mm}{5mm} \widetext \medskip }
\def\endwide{
        \hspace*{3.5in}~\rule[-5mm]{.1mm}{5mm}\rule{3.5in}{.1mm}
        \begin{multicols}{2}\narrowtext \vspace*{-1.0cm} \noindent }

\begin{document}
\bibliographystyle{unsrt}
\draft

\title{Persistent current in a mesoscopic ring with diffuse surface 
       scattering}

\author{K.V. Samokhin$^*$}

\address{Cavendish Laboratory, University of Cambridge, Madingley Road,
         Cambridge CB3 0HE, UK}
  
\date{\today}

\maketitle

\begin{abstract}
The persistent current in a clean mesoscopic ring with ballistic electron 
motion is calculated. The particle dynamics inside a ring is assumed to 
be chaotic due to scattering at the surface irregularities of atomic 
size. This allows one to use the so-called ``ballistic'' supersymmetric 
$\sigma$ model for calculation of the two-level correlation function in the 
presence of a nonzero magnetic flux.   
\end{abstract}


\begin{multicols}{2}\narrowtext 

Since the pioneering work of B\"uttiker, Imry and Landauer \cite{BIL83},  
persistent currents in small metallic rings have been a subject of 
great theoretical and experimental interest. While earlier observations were
made on disordered samples with diffusive electron dynamics \cite{exp},
more recent experiments measured the persistent currents in high mobility
semiconductor heterostructures with the elastic mean free path 
$l\simeq 1.3L$ ($L$ is the system size) \cite{MCB93}.
The semiclassical electron dynamics in such clean samples is ballistic and
can be either regular (integrable) or chaotic. The former possibility seems
to be rather exceptional, since any deviation of the sample shape from a
perfect annulus, however small, breaks down the integrability of a system. 
The purpose of this paper is to calculate the persistent current
in a ballistic ring, in which a bulk disorder is absent but the electron 
dynamics is nevertheless chaotic due to the multiple surface scattering. 
The electron-electron interaction is neglected. 

According to Ref. \cite{pers}, the average current for a canonical ensemble 
(i.e., for a fixed number of electrons in a sample) is given by the following
formula:
\begin{equation}
\label{pers_current_def}
 I=-\frac{s}{2\Delta}\frac{\partial}{\partial\Phi}\int d\epsilon_1
 d\epsilon_2\;n(\epsilon_1)n(\epsilon_2)K(\epsilon_1,\epsilon_2;\Phi)
\end{equation}
(we use the units in which $\hbar=c=1$).
Here $s$ is the spin degeneracy, $\Delta=(\rho_0 V)^{-1}$ is the mean level 
spacing in the system, $\rho_0$ is the average density of states, $V$ is the 
system volume, $n(\epsilon)$ is the Fermi distribution function, 
$K(\epsilon_1,\epsilon_2;\Phi)$ is the dimensionless two-level correlation 
function at a nonzero magnetic flux $\Phi$, which depends on the energy
difference $\epsilon_2-\epsilon_1=\omega$:
\begin{equation}
\label{level_corr_def}
 K(\omega,\Phi)=\frac{1}{\rho_0^2}\langle\delta\rho(E+\omega,\Phi)
 \delta\rho(E,\Phi)\rangle
\end{equation}  
($\delta\rho(E,\Phi)=\rho(E,\Phi)-\rho_0$ is the deviation of the 
one-particle density of states from its average value). We restrict our 
analysis to the limit of $T\gg\Delta$, where the formula 
(\ref{pers_current_def}) is valid. 

\begin{figure}
\begin{center}
\leavevmode
\epsfysize=0.7 \linewidth
\epsfbox{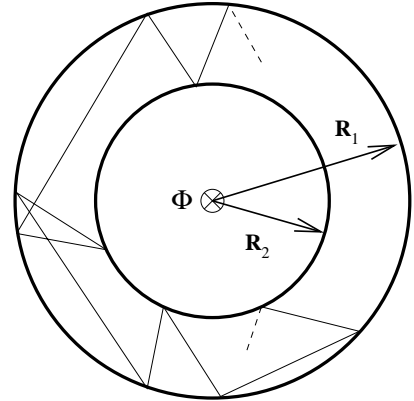}
\caption{The annular quantum billiard with diffusively scattering 
         surfaces ($\Phi$ is the magnetic flux). The light line shows a 
         typical trajectory of an electron inside the ring.}
\label{billiard}
\end{center}
\end{figure}

We assume that the sample has the shape of a planar coaxial ring with outer
and inner radii $R_1$ and $R_2$, respectively, threaded by a solenoid carrying 
a flux $\Phi$ (see Fig. \ref{billiard}). We consider here only narrow 
rings with $d=R_1-R_2\ll R=(R_1+R_2)/2$. For this geometry, 
the number of transverse channels $N=mv_{\rm F}d/2\pi$, the average 
density of states $\rho_0=m/2\pi$ and the mean level spacing 
$\Delta=(mRd)^{-1}$. 

In order to calculate the two-level correlation function 
(\ref{level_corr_def}), we use the supersymmetric $\sigma$ model, which has 
become a powerful tool in the theory of disordered metals \cite{Efetov} and
has been adapted to the description of the classically chaotic systems as well
\cite{MKh95,ASAA96}. Here it is appropriate to 
emphasize that in a clean ring, where a natural ensemble is absent but the 
dynamics is classically chaotic, the averaging in Eq. (\ref{level_corr_def}) 
must be performed over a wide energy band \cite{ASAA96}, 
in contrast to the case of a disordered ring, where the angular brackets in 
Eq. (\ref{level_corr_def}) 
imply averaging over different realizations of the random potential. 
If the shape of a sample is highly symmetric and its surface is smooth on 
the atomic scale, then the specular boundary conditions commonly 
used in chaotic billiards give rise to integrability of the system, and the 
whole approach based on using the supersymmetric $\sigma$ model fails 
(this case was studied in Ref. \cite{JRU96}, using the semiclassical trace 
formulas). 
In order to make the particle dynamics chaotic, one can, for instance, 
slightly deform the shape of the billiard to break its 
perfect rotational symmetry \cite{BR86,RM98}. However, in this approach, 
the level correlations can be calculated only numerically. 
Another way to achieve the chaotic regime, at the same time preserving 
the macroscopic symmetry of a sample, is to assume that each act of the 
surface reflection 
is stochastic itself, i.e., the incident particle gets reflected in some random
direction at the surface. This model is commonly referred to as the diffuse 
reflection and applies to the surfaces which are rough on the atomic scale, 
which seems to be quite reasonable physical assumption \cite{GE65}. After 
several reflections at the walls, the dynamics of an electron becomes fully 
chaotic.

At $\omega\gg\Delta$, the two-level correlation 
function (\ref{level_corr_def}) can be calculated perturbatively, using the 
``ballistic'' version of the supersymmetric $\sigma$ model generalized to 
the presence of a nonzero magnetic flux \cite{SAA97}:
\begin{equation}
\label{corr_fun}
 K(\omega,\Phi)=\frac{\Delta^2}{2\pi^2}\,{\rm Re}\sum\limits_i
 \frac{1}{[i\omega-\lambda_i(\Phi)]^2}
\end{equation}
where $\lambda_i(\Phi)$ are the eigenvalues of the (Cooperon) Liouville 
operator 
\begin{equation}
\label{L}
 {\cal L}=v_{\rm F}{\bf n}(\nabla_{\bf r}-2ie{\bf A})
\end{equation}
inside the Aharonov-Bohm billiard (${\bf n}$ is the direction of momentum). 
The region of small frequencies 
$\omega\ll\Delta$, where the perturbative approach fails and the level
correlations are described by the universal formulas of
the random matrix theory \cite{MP83}, lies beyond the limits of applicability 
of the thermodynamic approach to the description of persistent currents. 
The expression (\ref{corr_fun}) is a direct 
analog of the Altshuler-Shklovskii spectral function for diffusive systems 
\cite{AS86}. Rewriting the Fermi distribution functions in Eq.
(\ref{pers_current_def}) as Matsubara sums and integrating 
over energies, we end up with the following expression:
\begin{equation}
\label{I_Mats}
 I=-\frac{2\pi s}{\Delta} T\sum\limits_{n>0}\omega_n\frac{\partial 
 K(i\omega_n,\Phi)}{\partial\Phi},
\end{equation}
where $\omega_n=2\pi nT$. The sum over $n$ on the right-hand side is 
convergent due to the presence of the differentiation over flux. 

The spectrum of the Liouville operator is determined by the eigenvalue equation
\begin{equation}
\label{eigen_eq}
 v_{\rm F}{\bf n}(\nabla_{\bf r}-2ie{\bf A})f({\bf r},{\bf n})=\lambda
 f({\bf r},{\bf n}),
\end{equation}
with some boundary conditions at the surfaces of the ring. 
Due to the similarity of Eq. (\ref{eigen_eq}) to the 
Boltzmann kinetic equation, $f({\bf r},{\bf n})$ having the
meaning of the classical distribution function, the boundary conditions at the 
diffusely reflecting surface ${\bf r}={\bf R}$ with the outward normal
${\bf N}$ can be imposed by analogy with the classical kinetic theory.  
The distribution function of reflected particles can be represented as 
$f({\bf R},{\bf n})=pf_0({\bf R})+(1-p)f({\bf R},\bar{\bf n})$ 
\cite{Fuchs38}, where $0\leq p\leq 1$ is ``the diffuseness coefficient'', 
$f_0$ is an isotropic distribution function, and 
$\bar{\bf n}={\bf n}-2({\bf n}{\bf N}){\bf N}$ is the direction of
specular reflection. In this paper, we consider an
isotropic diffuse scattering with $p=1$, corresponding to the limit of
``strong chaos'' (for the discussion of applicability of this model to real
experimental samples, see below) and the boundary condition, which follows 
from the particle number conservation, takes the form
\begin{equation}
\label{Fuchs}
 \frac{1}{\pi}f({\bf R}_i,{\bf n})|_{({\bf n}{\bf N}_i)<0}=\int_{({\bf n}' 
  {\bf N}_i)>0} d{\bf n}'\;({\bf n}'{\bf N}_i)f({\bf R}_i,{\bf n}').
\end{equation}
Here $\int d{\bf n}=\int d\phi/2\pi$ ($\phi$ is the angle between the 
direction of momentum of incident particles and the outward normal ${\bf N}$),
and $i=1,2$ correspond to the outer and inner surfaces of the ring. 
The distribution function of reflected particles on the left-hand side of 
Eq. (\ref{Fuchs}) does not depend on ${\bf n}$. A similar approach was used 
in Ref. \cite{disc} for calculation of the corrections to the universal 
level correlations in a two-dimensional disk without magnetic flux.

\begin{figure}
\begin{center}
\leavevmode
\epsfysize=0.7 \linewidth
\epsfbox{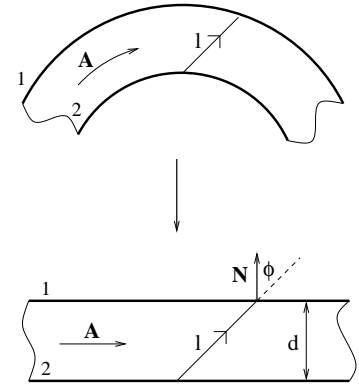}
\caption{The transformation of a ring into a narrow strip of width $d=R_1-R_2$
         and length $L=2\pi R=\pi(R_1+R_2)$. $A=\Phi/2\pi R$ is the vector 
         potential created by the solenoid.}
\label{geometry}
\end{center}
\end{figure}

Since the magnetic field is absent inside the ring, the trajectory of
a particle between collisions with the walls is a straight line. 
Equation (\ref{eigen_eq}) can be solved along the trajectory
\begin{equation}
\label{gen_solution}
 f(l)=f(0)\exp\left(\frac{\lambda}{v_{\rm F}}l\right)
 \exp\left(2ie\int_0^l{\bf A}d{\bf l}\right).
\end{equation}
This expression allows one to establish a relation between $f({\bf R}_1,
{\bf n})|_{({\bf n}{\bf N}_1)>0}$ and $f({\bf R}_1,{\bf n})|_{({\bf n}
{\bf N}_1)<0}$ or $f({\bf R}_2,{\bf n})|_{({\bf n}{\bf N}_2)<0}$, and also
between $f({\bf R}_2,{\bf n})|_{({\bf n}{\bf N}_2)>0}$ and 
$f({\bf R}_1,{\bf n})|_{({\bf n}{\bf N}_1)<0}$ in Eq. (\ref{Fuchs}) 
and obtain a rather cumbersome algebraic equation for the 
eigenvalues of the Liouville operator 
in a ring of arbitrary width. Fortunately, in the case of a narrow ring, the 
problem can be considerably simplified, since we can replace our annular 
billiard by a strip of length $L=2\pi R$ and width $d$ such that 
$\delta=d/L\ll 1$ (see Fig. \ref{geometry}). In addition, the vector 
potential can be put constant inside the sample: $A_{\theta}=\Phi/L$
($\theta$ is the polar angle in real space). 
This simplification implies that the contribution from the trajectories 
connecting two points at the outer wall is neglected (it can be checked that 
this contribution is indeed small at $\delta\ll 1$). However, there is an 
important property of the annular geometry which should be taken into account, 
namely the finiteness of the flight length $l(\phi)$ between successive
collisions with the walls ($l\ll L$). This feature can be restored in the 
strip billiard if to assume that there exists the maximum scattering 
angle $\phi_0$ such that $\sin\phi_0=R_2/R_1\simeq 1-2\pi\delta$. 

Let $x=R\theta$, then we obtain, from Eq. (\ref{gen_solution}) and Fig. 
\ref{geometry}:
\begin{equation}
\label{f12}
\left. \begin{array}{rcl}
 &&f_{1,>}(x,\phi)=f_{2,<}(x-d\tan\phi)\\
 &&\qquad\qquad\times\exp\left(\frac{\lambda}{v_{\rm F}}\frac{d}{\cos\phi}
  \right)\exp\left(2\pi i\frac{2\Phi}{\Phi_0}\frac{d\tan\phi}{L}\right)\\
 &&f_{2,>}(x,\phi)=f_{1,<}(x+d\tan\phi)\\
 &&\qquad\qquad\times\exp\left(\frac{\lambda}{v_{\rm F}}\frac{d}{\cos\phi}
  \right)\exp\left(-2\pi i\frac{2\Phi}{\Phi_0}\frac{d\tan\phi}{L}\right).
 \end{array} \right.
\end{equation}
Here $f_{i,>(<)}(x,\phi)=f({\bf R}_i,{\bf n})|_{({\bf n}{\bf N}_i)>0(<0)}$ and
$\Phi_0=2\pi\hbar c/e$ is the flux quantum. 
It is convenient to expand the functions $f_i$ in the Fourier series: 
$f_{i,<}(x)=\sum_q f_{i,q}e^{iqx}$. Since $f(x+L,{\bf n})=f(x,{\bf n})$, 
the wave number is quantized: $q=m(2\pi/L)$, where $m=0,\pm 1,...$. 
Substitution of the expressions (\ref{f12}) in Eq. (\ref{Fuchs}) results in 
the following equation for the eigenvalues $\lambda_{m,k}(\Phi)=
(v_{\rm F}/d)z_{m,k}(\Phi)$ of the Liouville operator (\ref{L}):
\begin{equation}
\label{gen_eq}
  F_m(z,\nu)\equiv A_m^2(z,\nu)-1=0.
\end{equation}
Here $\nu=2\Phi/\Phi_0$ and
\begin{eqnarray}
\label{A}
 A_m=\frac{1}{\sin\phi_0}\int_0^{\phi_0}d\phi\;\cos\phi\exp(z\sec\phi)
  \nonumber\\
 \times\cos[2\pi\delta(m-\nu)\tan\phi].
\end{eqnarray}
To guarantee correct normalization of the distribution function of 
reflected particles, the prefactor $\sin^{-1}\phi_0$ has been included 
in the definition of $A_m$.

The eigenvalues of the Liouville operator at fixed $m$ are labeled by 
$k=0,1,2,...$, and have complex conjugated partners for all $(m,k)$. 
At $\Phi=0$, Eq. (\ref{gen_eq}) has the solution $z_{0,0}=0$ 
corresponding to the equilibrium distribution function, which does not 
depend on ${\bf n}$ and ${\bf r}$. It is clear from Eq. (\ref{A}) that the 
eigenvalues $\lambda_{m,k}(\Phi)$ have the following property: 
$\lambda_{m,k}(\Phi+\Phi_0/2)=\lambda_{m+1,k}(\Phi)$ 
so that any quantity which can be represented in the form 
$\sum_{m,k}h(\lambda_{m,k}(\Phi))$, where $h(\lambda)$ is some 
function, must be periodic function of $\Phi$ with the period of half the 
flux quantum $\Phi_0$. Note that, since the eigenvalues of the Liouville 
operator physically correspond to the relaxation rates of different harmonics 
of a nonequilibrium classical distribution function, the real parts of all 
$\lambda_{m,k}$ are positive.

Using the representation of the sum over $i=(m,k)$ in Eqs. (\ref{corr_fun}) 
and (\ref{I_Mats}) as an integral over a contour $C$ enclosing all zeros of 
the function $F_m(\nu,z)$ in the complex plane of $z$:
\begin{eqnarray*}
 &&\sum\limits_{m,k}\frac{1}{[\omega_n+\lambda_{m,k}(\Phi)]^2}\\
 &&\hspace*{0.5cm}=\left(\frac{d}{v_{\rm F}}\right)^2\sum\limits_m
 \oint_C\frac{dz}{2\pi i}\frac{1}{(z+\omega_nd/v_{\rm F})^2}\frac{\partial
 }{\partial z}\ln F_m(z,\nu)\\
 &&\hspace*{0.5cm}=-\left(\frac{d}{v_{\rm F}}\right)^2\sum\limits_m
 \left.\frac{\partial^2}{\partial z^2}\ln F_m(z,\nu)\right|_{z=-
 \omega_n d/v_{\rm F}},
\end{eqnarray*}
we finally obtain 
\begin{eqnarray}
\label{result}
 \frac{I}{I_0}&=&\frac{2s\delta^2}{\pi N^3}\left(\frac{T}{\Delta}\right)^2 
   {\rm Re}\sum\limits_{n>0}\sum\limits_m n \nonumber \\
 &&\times\left.\frac{\partial}{\partial\nu}\frac{\partial^2}{\partial z^2}
 \ln F_m(z,\nu)\right|_{z=-2\pi\delta nT/N\Delta},
\end{eqnarray}
where $I_0=ev_{\rm F}/L$ is the current carried by a single electron state in
an ideal one-dimensional ring, and $F_m(z,\nu)$ is given by Eq. 
(\ref{gen_eq}). 

Due to the existence of different energy scales in the system, the 
temperature dependence of the persistent current is characterized by several 
distinct regimes. The smallest energy scale is the mean level 
spacing $\Delta$, which also limits the applicability of the thermodynamic 
approach itself. Two other scales are given by the inverse times $t_L^{-1}
=v_{\rm F}/L=N\Delta$ and $t_d^{-1}=v_{\rm F}/d=N\Delta/\delta$. 
It follows from Eqs. (\ref{result}) and (\ref{A}) that
at $T\gg t_d^{-1}$ the persistent current is exponentially small: $I\sim I_0
\exp(-T/N\Delta)$. 

In a multichannel ring, there also exists yet another energy scale 
$\Delta\ll E_c\ll N\Delta$, 
whose origin can be most easily understood if to return to Eq. (\ref{I_Mats}).
Using the identity $a^{-2}=\int_0^\infty dye^{-ay}y$ and calculating the
sum over $n$, we end up with the following expression
\begin{equation}
\label{I_varphi}
 \frac{I}{I_0}=\frac{s}{2\pi^4\delta}\frac{\Delta}{T}\int_0^\infty
 \frac{d\xi\;\xi^2}{\sinh^2\xi}\,\varphi(\xi;T,\Phi),
\end{equation}
where 
\begin{equation}
\label{varphi}
 \varphi=\frac{\Phi_0d}{2v_{\rm F}}\,{\rm Re}\sum\limits_{m,k}
 \frac{d\lambda_{m,k}(\Phi)}{d\Phi}\exp\left(-\frac{\lambda_{m,k}
 (\Phi)\xi}{\pi T}\right).
\end{equation}
At sufficiently low temperatures, the dominant contribution comes from 
$\lambda_{0,0}(\Phi)$, which is the solution of Eq. (\ref{gen_eq}) 
with smallest real part. At $\Phi\to 0$, the energy $E_c$ coincides with
$\lambda_{0,1}(\Phi=0)$, and at $T\ll E_c$ the sum over 
$(m,k)$ in Eq. (\ref{varphi}) can be replaced by its value at $m=k=0$.  
It follows from Eq. (\ref{gen_eq}) that $z_{0,0}(\nu)=2\pi\delta^2|\ln\delta|
\nu^2$ at $\nu\to 0$, and we obtain:
\begin{equation}
\label{Curie}
 \frac{I}{I_0}=\frac{2s}{3\pi}\left(\delta\ln\frac{1}{\delta}\right)
 \frac{\Delta}{T}\frac{\Phi}{\Phi_0}.
\end{equation}
Thus the persistent current and also the orbital magnetic moment $M=\pi R^2I$
of a ballistic ring exhibit a Curie-type response on magnetic flux
at $\Delta\ll T\ll E_c$. 

The energy $E_c$ is associated with the relaxation time $t_c=E_c^{-1}$ of any
initially nonequilibrium state of the system to a spatially uniform 
and isotropic distribution (the ergodic time). In the systems with diffusive 
electron dynamics, $t_c=L^2/D$ ($L$ is the system size, $D$ is the diffusion 
coefficient) and $E_c$ is called Thouless energy. In general, 
it is not so easy, however, to say what the ergodic time is in ballistic 
systems \cite{AGM96}. In our case, it can be estimated from simple 
physical considerations, without solving Eq. (\ref{gen_eq}) explicitly, as 
follows. The typical flight time $t_f$ between two reflections (see Fig. 
\ref{geometry}) is
\begin{equation}
\label{t_f}
 t_f=\frac{1}{v_{\rm F}}\sqrt{\langle l^2(\phi)\rangle}\sim t_L\delta^{3/4}.
\end{equation}
Since the particle completely ``forgets'' the initial direction of its 
momentum after each collision with the walls, the characteristic time 
for establishing an isotropic momentum distribution is of the order of 
$t_f$. In contrast, the characteristic time of filling uniformly all 
available configuration space is much longer. Indeed, the typical
displacement between two collisions is given by $\sqrt{\langle(\delta x)^2
\rangle}\simeq v_{\rm F}t_f\sim L\delta^{3/4}$. According to the central 
limit theorem, the probability to find a particle at the distance $\Delta x=
\sum_{i=1}^M\delta x_i$ after $M\gg 1$ collisions obeys
the Gaussian distribution with the standard deviation $\sqrt{\langle
(\Delta x)^2\rangle}=\sqrt{M\langle(\delta x)^2\rangle}$. Substituting here 
$M\simeq t/t_f$ and using Eq. (\ref{t_f}), we obtain
\begin{equation}
 \langle(\Delta x(t))^2\rangle=D_{\rm eff}t,\qquad 
  D_{\rm eff}\sim v_{\rm F}^2t_f.
\end{equation}
Thus the motion of an electron along the circumference of the ring is 
in fact diffusive at $t\gg t_f$, and
\begin{equation}
 E_c=\frac{D_{\rm eff}}{L^2}\sim t_L^{-1}\delta^{3/4}.
\end{equation}
In a narrow multichannel ring the hierarchy of the characteristic energy 
scales looks as follows: $\Delta\ll E_c\ll t_L^{-1}\ll t_f^{-1}\ll 
t_d^{-1}$. Note that all our results are valid if one can neglect the 
deflection of the electron trajectories inside the ring by an external
magnetic field, which is the case if the inverse cyclotron radius $R_c^{-1}=
(eB/mv_{\rm F})$ is smaller than the inverse typical flight length $(v_{\rm F}
t_f)^{-1}$. This condition, rewritten as $\Phi/\Phi_0\ll N\delta^{-7/4}$,
is always satisfied in narrow rings. 
 
In order to facilitate comparison of our results with the experimental data, 
let us rewrite Eq. (\ref{Curie}) in a different form, using the identity
$\Delta/T=N^{-1}(L_T/L)$, where $L_T=v_{\rm F}/T$ is the length scale 
associated with temperature. In the experiment of Ref. \cite{MCB93}, 
$\delta\simeq 0.1$, $N\simeq 4$, $L_T/L\simeq 5$, so that 
$T{<\atop\sim}\Delta$. Due to the presence of the 
factor $\delta\ln\delta\ll 1 $ in Eq. (\ref{Curie}), the predicted current 
turns out to be smaller than the experimentally observed (and also than the
theoretically calculated for the case of specular reflection \cite{JRU96}). 
This discrepancy can be attributed to the fact that, because of the low 
density of carriers, the Fermi wavelength greatly exceeds the size 
of the surface irregularities, so that a considerable fraction of 
particles gets reflected specularly rather than diffusely (i.e., $p<1$), 
and one should describe the semiclassical dynamics in the experimental
conditions of Ref. \cite{MCB93} as ``weakly chaotic''. 

To summarize, we calculated the persistent current in a small clean metal 
ring, in which the electron dynamics is chaotic due to the stochastic surface 
scattering. A general analytical expression for the
persistent current is derived in the limit of ``strong chaos'', and a
Curie-type orbital magnetic response on a small external flux is predicted at 
$\Delta\ll T\ll E_c$. 

\bigskip

The author would like to thank B. Simons for interest in this work and 
useful discussions. 
This work was financially supported by the Engineering and Physical Sciences 
Research Council, UK.

\end{multicols}

\end{document}